\documentclass[aps,prx,twocolumn,superscriptaddress,longbibliography,floatfix,10pt]{revtex4-2}
\usepackage{graphicx}
\usepackage{amsmath}
\usepackage{xcolor}
\usepackage{hyperref}
\usepackage{orcidlink}

\begin{document}

\title{Superconducting Qubit Decoherence Correlated with Detected Radiation Events}

\author{Alessandro R. Castelli\,\orcidlink{0000-0002-5202-9839}}
\email{castelli1@llnl.gov}
\affiliation{Lawrence Livermore National Laboratory, Livermore, CA, USA}
\author{Kristin M. Beck\,\orcidlink{0000-0003-2486-4164}}
\affiliation{Lawrence Livermore National Laboratory, Livermore, CA, USA}
\author{Loren D. H. Alegria\,\orcidlink{0000-0001-8960-2937}}
\affiliation{Lawrence Livermore National Laboratory, Livermore, CA, USA}
\author{Luis A. Martinez\,\orcidlink{0000-0002-2931-0833}}
\affiliation{Lawrence Livermore National Laboratory, Livermore, CA, USA}
\author{Kevin R. Chaves\,\orcidlink{0000-0001-9542-0275}}
\affiliation{Lawrence Livermore National Laboratory, Livermore, CA, USA}
\author{Sean R. O’Kelley\,\orcidlink{0000-0003-0711-6471}}
\affiliation{Lawrence Livermore National Laboratory, Livermore, CA, USA}
\author{Nicholas Materise\,\orcidlink{0000-0002-7353-4770}}
\affiliation{Lawrence Livermore National Laboratory, Livermore, CA, USA}
\author{Jonathan L DuBois\,\orcidlink{0000-0003-3154-4273}}
\affiliation{Lawrence Livermore National Laboratory, Livermore, CA, USA}
\author{Yaniv J. Rosen\,\orcidlink{0000-0003-4671-2305}}
\affiliation{Lawrence Livermore National Laboratory, Livermore, CA, USA}

\begin{abstract}
Most quantum error correction (QEC) protocols for superconducting qubits assume spatially and temporally uncorrelated decoherence events; however, recent evidence suggests that cosmic radiation induces spatially correlated errors. We present a platform that sandwiches a superconducting transmon qubit between two microwave kinetic inductance detector (MKID) arrays, enabling real-time detection of radiation-induced phonon bursts. By synchronizing MKID event detection with single-shot measurements of qubit energy relaxation ($T_1$) and phase coherence ($T_2$), we observe statistically significant reductions in both $T_1$ and $T_2$---up to 30.5\%---immediately following dual MKID events attributed to penetrating muons. Our findings directly link radiating events to correlated qubit decoherence. Furthermore, our experimental platform provides a foundation for systematic studies of radiation effects, the development of shielding and mitigation techniques, and the refinement of error-correction algorithms tailored to correlated noise sources.
\end{abstract}

\maketitle

\section{Introduction}

Current superconducting quantum processing units (QPUs) contain tens of physical qubits on a single silicon or sapphire chip~\cite{Kim2023,Hoke2023,Maciejewski2024}. These individual qubits decohere by exchanging energy or through interactions that scramble their phase. To correct these physical errors and operate increasingly large qubit architectures, quantum error-correction (QEC) algorithms have been proposed and demonstrated in proof-of-principle experiments with tens of qubits~\cite{Chiaverini2004,Schindler2011,Roffe2019,Google2023}.

Radiation is known to create spatially correlated disruptions in superconducting and other condensed matter qubits~\cite{Wilen2021,Vepsalainen2020,Thorbeck2023,Tennant2022,Gordon2022}. Penetrating radiation (muons) and other local sources (gammas) can deposit a large amount of energy relative to the superconducting gap~\cite{Cardani2023}. As QPUs become physically larger, they become more susceptible to phonon bursts caused by radiation due to their larger cross-sectional area. These radiation events cause spatially correlated errors that are not protected against in current QEC algorithms~\cite{Vischi2022,Oliveira2023}.

While radiation effects have been widely observed in qubits, the observed effects differ. Some groups report catastrophic spontaneous energy decay, in which qubits have reduced energy coherence ($T_1$) over many milliseconds~\cite{McEwen2022,Harrington2025}. Others report that charge-parity jumps are the primary effect, with bit flips ($T_1$ decay) occurring at high quasiparticle density~\cite{Li2025}. However, it is difficult to directly evaluate the effect of high energy events on the qubits. To address this issue, we use microwave kinetic inductance detectors (MKIDs) to monitor radiation interactions while simultaneously performing single shot measurements on qubits.

MKIDs are mature photon detectors that have been utilized in telescopes, X-ray spectrometers, and particle physics detectors for neutrinos and dark matter~\cite{Ulbricht2021}. They can detect electromagnetic energy ranging in frequency from X-ray to far infrared and are sensitive to any event that breaks Cooper pairs. Aluminum MKIDs are sensitive above approximately $3.6 \times 10^{-4}$~eV~\cite{Day2003,Fruitwala2020,Du2022}.

Gamma rays, muons, and other radioactive particles eject electrons from tightly bound shells in the substrate and create large-energy phonons. Depending on their energies, these phonons cascade through the substrate and downconvert into lower energy phonons~\cite{Martinis2021}. If these phonons hit the superconducting layer, they excite Cooper pairs into quasiparticles and trigger a detectable event in one or more MKIDs~\cite{Mannila2022}.

In this work, we first introduce our platform for measuring correlated qubit errors with radiation events. A superconducting qubit sample is placed between two MKID arrays. We then describe our method for reading out these sensitive radiation detectors and demonstrate that we can measure the energy deposited from a radiation event in each array as well as distinguish between penetrating and nonpenetrating radiation. Finally, we show a correlation between penetrating radiation events and reduced qubit coherence.

\section{Methodology}

\begin{figure}[ht]
    \centering
    \includegraphics[width=1\linewidth]{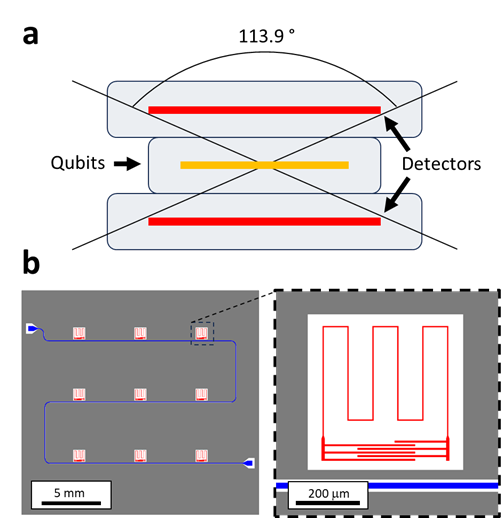}
    \caption{\textbf{Experimental platform for correlated radiation event detection.} (a) Schematic showing a superconducting qubit chip sandwiched between two microwave kinetic inductance detector (MKID) arrays. This geometry enables direct evaluation of correlations between qubit performance and incident penetrating radiation. The MKID array's opening angle ($113.9^\circ$) provides coverage of the expected muon flux distribution. Each chip is housed in a gold-coated copper box to ensure thermalization and microwave shielding, and all three layers are fixed to a common mounting bracket at the dilution refrigerator mixing chamber stage. (b) Circuit layouts for the top and bottom MKID arrays, each comprising a $3 \times 3$ grid of individual MKIDs (shown in red). Each MKID has a unique resonance frequency, adjusted by the arm length of the interdigitated capacitor. All MKIDs are read out via capacitive coupling to a shared $50~\Omega$ impedance transmission line (shown in blue), enabling simultaneous monitoring of the entire array via a single microwave feedline.}
    \label{fig:setup}
\end{figure}

We detect correlated high-energy events using three layers of devices (see Fig.~\ref{fig:setup}). The three devices are comprised of two MKID detector array chips with a 6-qubit device chip mounted between the top and bottom layers (see Appendix~\ref{app:setup}). In this way, any penetrating radiative events detected by the qubit layer are more likely to be detected by the top and bottom layers as well. The whole device is mounted such that each chip is perpendicular to the $0^\circ$ zenith angle vector so that incident muons have the largest cross-sectional area target possible. Each chip is mounted in its own gold-plated copper sample box and secured with a thermally anchoring adhesive varnish. The sample boxes are designed to mount to each other in a three-layer configuration and subsequently to a copper mounting bracket that easily attaches to a sample holder at the mixing chamber stage of a dilution refrigerator.

The samples and sample holders were designed to maximize the solid angle for the detectors while maintaining optimal device operation and reliability. The achieved solid angle accounts for incident muons whose trajectories lie within an angular range of $\pm 57^\circ$. Outside of this range, the expected muon count falls below $100~\text{counts/hour/m}^2$ ($3 \times 10^{-6}~\text{counts/s/cm}^2$)~\cite{workman2022prog}. The top and bottom detector chips are placed symmetrically above and below the qubit chip, which is oriented such that the center of the chip is located at the apex of the solid angle (see Fig.~\ref{fig:setup}). This gives the highest likelihood of having muons pass through both detector substrates.

The MKID arrays are fabricated on a $20 \times 20 \times 0.65~\text{mm}$ sapphire substrate with a $20~\text{nm}$ film of aluminum. These nine devices are connected to a single transmission line that is used to monitor their resonant frequencies. Each MKID is constructed as a lumped element device with an inductive element and an interdigitated capacitor. The frequency of each MKID is adjusted by slightly increasing the capacitance in each of the nine individual detectors on the chip. The device resonances were designed to span approximately $200~\text{MHz}$ so that they would be well within the bandwidth of a single channel of a $1~\text{Gs/s}$ digital-to-analog converter.

Separate measurement chains connect each individual chip to the room temperature control system. The isolation between these measurement chains is more than $50~\text{dB}$ to prevent inadvertent crosstalk. Attenuation and filtering between the room temperature environment and the $10~\text{mK}$ mixing chamber reduce the number of thermal photons present at the frequencies of interest. Amplifiers on the return path are used to increase the signal-to-noise ratio~\cite{Wu2020}.

A room-temperature microwave control system is connected to the dilution refrigerator wiring for the purpose of driving each array of MKIDs and the qubit sample. The microwave control system is a Quantum Machines Operator X+ (OPX+) controller. The OPX+ has 10 analog output lines with $\pm 350~\text{MHz}$ bandwidth, $\pm 0.5~\text{V}$ range, and 16-bit resolution. To generate control signals at the required GHz-range frequencies, pairs of output lines are upconverted using a single sideband mixer and static local oscillators. Each OPX+ has two analog input lines with similar specifications: $\pm 350~\text{MHz}$ bandwidth, $\pm 0.5~\text{V}$ range, and 12-bit resolution. Return signals are downconverted into this bandwidth using the same static oscillator used for upconversion.

This control system is used to continually monitor the MKID arrays’ resonance frequency. An MKID’s sensitivity stems from its large kinetic inductance fraction. Impinging radiation creates quasiparticles in the MKIDs’ superconducting film, which manifests as an increase in the film’s kinetic inductance. This increase shifts the resonator’s frequency lower by an amount well above the resonator’s noise levels. We can therefore probe the resonator after a radiation event and measure the frequency shift of each MKID affected. We continuously monitor the MKIDs and only record data when a change in phase and amplitude is detected. We record this buffer (approximately $15~\mu\text{s}$) and a subsequent detection window (approximately $700~\mu\text{s}$) for analysis. Recording triggered data enables long acquisition times ($>12~\text{hours}$) with reasonable size datasets ($<6~\text{Gb}$). We measure the MKID frequency in a pulsed configuration. This measurement scheme gives us better signal quality than a continuous measurement due to the specifics of our readout chain, the quality factor, and the nonlinearity of our devices.

\begin{figure}[ht]
    \centering
    \includegraphics[width=1\linewidth]{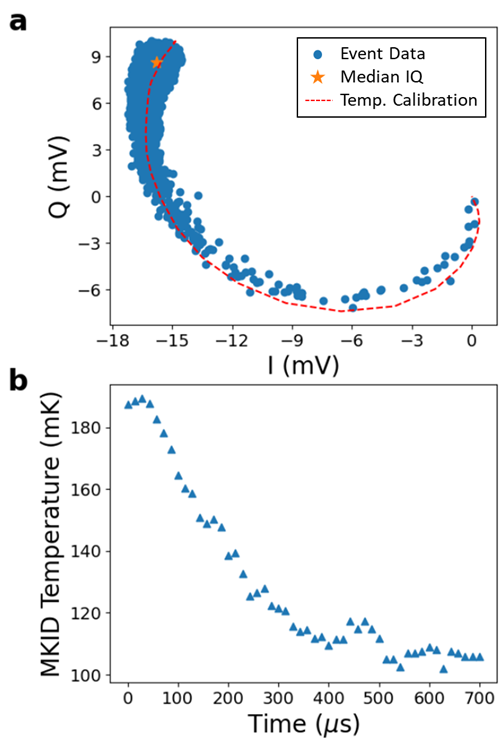}
    \caption{\textbf{MKID response to radiation events.} (a) Cumulative in-phase (I) and quadrature (Q) data for a single MKID, recorded over 12 hours with acquisition triggered when an event was detected on any MKID in the array. As the local temperature rises, the increased quasiparticle population broadens and shifts the resonance frequency, causing the signal amplitude to decrease accordingly. The resulting traces form a characteristic arc in IQ space that terminates near the origin. (b) Time-resolved, temperature-calibrated response of the same MKID to a single event, illustrating the relaxation of the resonator as quasiparticles recombine into Cooper pairs. Temperature calibration enables quantification of the energy deposited by the event.}
    \label{fig:IQ}
\end{figure}

\section{Results}

We track changes in individual MKID frequency and amplitude through the measured phase of the readout signal as a function of time. To increase the signal for our resonator’s quality factor and response, we use a ringdown measurement. In this measurement, we apply a pulse, typically approximately $4~\mu\text{s}$, which increases the energy in the resonator. Our readout begins after stopping the applied pulse, at which point we measure the energy leaving the resonator. Because we probe the resonators using ringdown measurements, the resulting plot in IQ space generally appears as a sweeping arc spiraling towards the origin of the IQ plane (see Fig.~\ref{fig:IQ}a), but the exact shape is slightly different for each resonator. Measurements are set to begin taking data when an event is detected and to return to standby after $700~\mu\text{s}$. Event detection counts as a shift in any detector’s I or Q signal larger than the threshold value.

When an event is detected on one MKID, the data for all MKIDs are recorded to look for correlated events. Using calibrated temperature data (see Appendix~\ref{app:calibration}), we can directly map the phase response of detected events to a temperature value (see Fig.~\ref{fig:IQ}b). However, the meaningful temperature range is limited to $100$-$300~\text{mK}$. This limitation exists because the detectors have minimal response below $100~\text{mK}$ and the detector signal arcs terminate at the origin of the IQ plane for temperatures above $300~\text{mK}$. The detector response is nonlinear, resulting in each resonator experiencing relatively larger frequency shifts and lower $Q$ values during more energetic events.

\begin{figure}[htbp]
    \centering
    \includegraphics[width=1\linewidth]{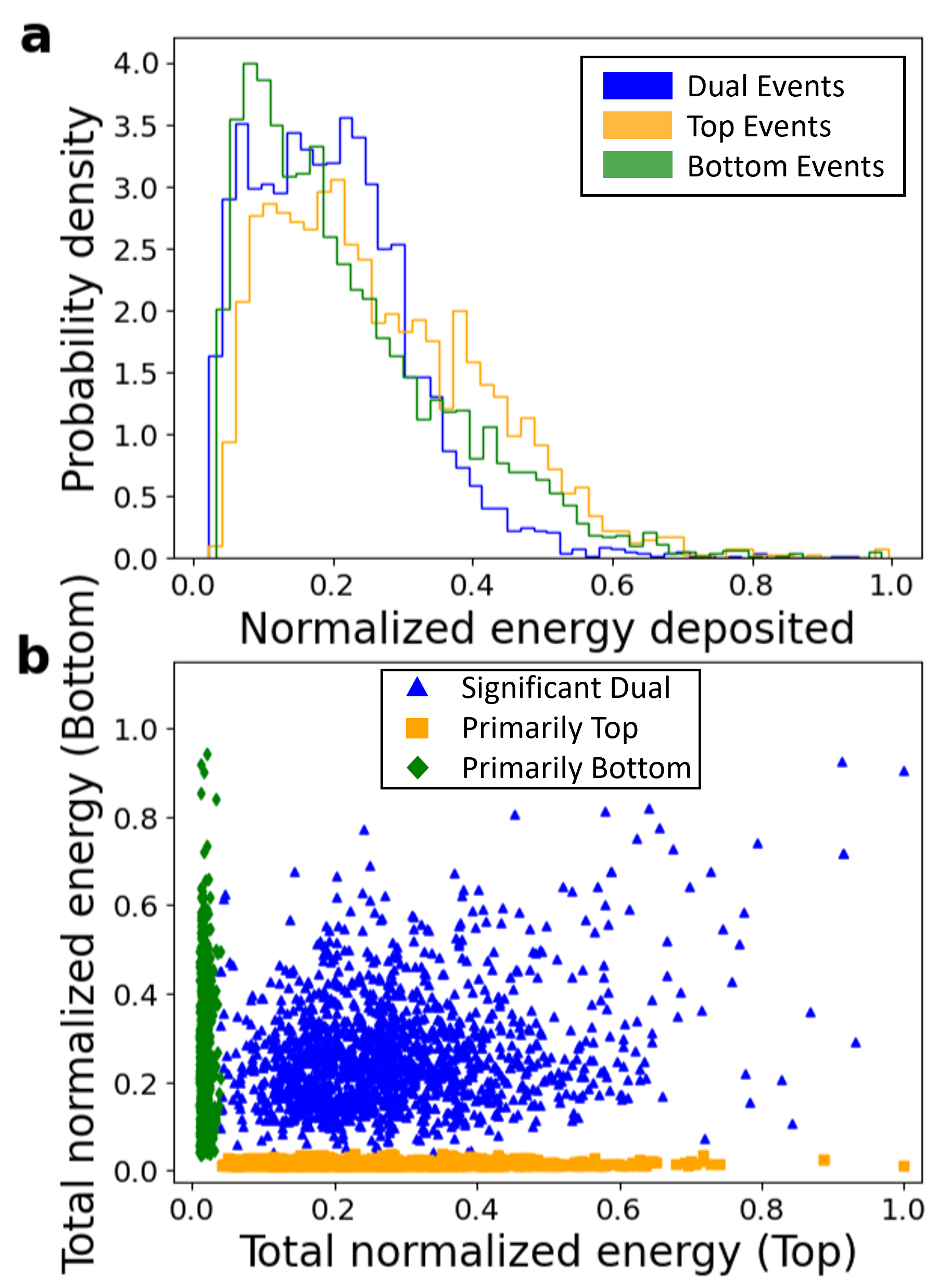}
    \caption{
        \textbf{MKID event energy distributions.}
        (a) Probability density of the normalized energy deposited for ‘top,’ ‘bottom,’ and ‘dual’ events. The energy tail for dual-detector events is shorter than for top-only and bottom-only events, likely reflecting differences in the type of radiation involved. Dual-detector events typically arise from penetrating particles, which deposit less energy per detector on average, whereas top-only and bottom-only events are more likely caused by localized sources that have shorter stopping distances, resulting in greater energy deposition.
        (b) Scatter plot of total normalized energy in dual-detector events, where at least one MKID in both the top and bottom arrays registers an event. A total normalized energy of 1 indicates that every detector on a substrate recorded the maximum detectable energy. The distribution reveals three event types: those where energy is deposited primarily in the top array (orange), primarily in the bottom array (green), and distributed between both arrays (blue).
    }
    \label{fig:fig3}
\end{figure}

We can use the temperature normalization to analyze the results of a long data acquisition period and infer more information about the radiation events occurring during that time. When comparing the histograms for dual-detector and single-detector events (see Fig.~\ref{fig:fig3}a), we see that the events deposit different energies. Specifically, dual detector events appear to deposit less energy on average per detector than single detector events. This is, at least in part, due to the large number of dual-detector events that register a large impact on one array and small impact on the other. The higher energy tail of the single-detection events could come from particles (gammas, alphas) with short stopping distances or particles with highly oblique impact paths relative to the detectors. These particles have a longer interaction distance and would deposit more energy than particles that travel perpendicular to the detectors (muons). The differences between the distributions for top-only and bottom-only detection events are most likely due to small differences in the relative sensitivity and noise floor of the top and bottom detectors.

Next, we look at the energy distribution in dual-detector events. In Fig.~\ref{fig:fig3}b, we see the total energy deposited in each detector for each dual detector event. We normalize each event to the total temperature range of that MKID and then average across the MKIDs on a substrate. A reading of 1 signifies that every detector on the substrate registered the maximum of the detectable temperature range. The distinct difference in the distribution of dual-detector events may be an indication of differences in energy distribution of penetrating and nonpenetrating events and suggest that there may be roughly two classes of penetrating events (weak and strong).

Dual-detector events account for approximately $25\%$ of all events and are most likely the result of cosmogenic muons passing through both layers. The most common event produces a detectable signal on all nine MKIDs on a chip, albeit with varying amounts of energy. When one chip sees events on all nine MKIDs, it is likely that the second chip will also see all nine. Dual events can potentially allow us to infer the trajectory of the radiation event by correlating resonator detection timing across both detector chips.

\begin{figure}[htbp]
    \centering
    \includegraphics[width=1\linewidth]{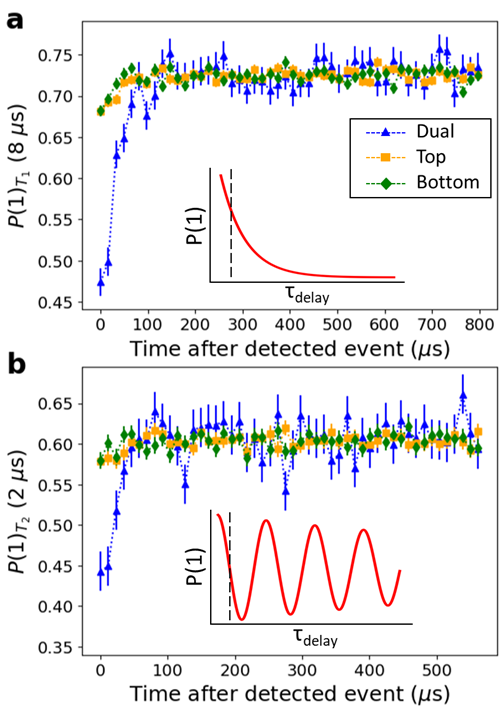}
    \caption{
        \textbf{Qubit coherence following MKID-detected events.} Mean expectation value of single shot (a) $T_1$ and (b) $T_2$ measurements as a function of time after event start, separated by event class. Insets in each panel illustrate the delay time used for measurement, with a schematic of the average qubit population expected from the pulse sequence. Qubit measurements triggered by dual-detector events (blue triangles) exhibit statistically significant changes in both coherence times. The smaller effects observed for top-only (orange squares) and bottom-only (green diamonds) events are likely due to incomplete detection of penetrating events (see discussion). Error bars represent $\pm1$ standard deviation.
    }
    \label{fig:fig4}
\end{figure}

When we began measuring qubits, we also increased MKID pixel number from a three-by-three array to a five-by-five array to achieve better spatial resolution. Experiments involving the entire three-chip setup demonstrate a correlation between dual-detector events and qubit decay and coherence times, but no correlation is observed during single (top or bottom only) MKID detections (see Fig.~\ref{fig:fig4}). The data for these experiments is filtered into three categories to illustrate the difference in the qubit’s $T_1$ and $T_2$ times for varying classes of detected events. The three classes include `dual' detection events, in which at least a single MKID on both chips register an event, as well as `top' and `bottom' detection events, in which only MKIDs on a singular chip registered an event.

We measured single shot $T_1$ and $T_2$ excited state population by picking a single delay time from $T_1$ and $T_2$ decay curves. We used a conditional reset pulse to drive the qubit into its ground state after every measurement to reduce data acquisition latency. After an event is detected, fifty complete MKID and qubit data acquisitions are recorded with a latency of approximately $10$-$15~\mu\text{s}$ depending on the measurement. The experimental details, including pulse sequencing, are given in further detail in Appendix~A. The acquired data captures the time-resolved event relaxation which lasts on the order of $100~\mu\text{s}$.

The $T_1$ measurements are performed in a single shot consisting of the application of a $\pi$-pulse that excites the qubit to the excited state, a set wait time (in this case $33\%$ of the qubit’s total $T_1$ time), and finally a readout pulse to acquire the qubit state. For the $T_2$ measurement, the qubit is first excited with a detuned Ramsey pulse which takes it to the equator of the Bloch sphere. We wait one quarter period, approximately $2~\mu\text{s}$, for the detuning to cause the qubit state to precess, then play another detuned Ramsey pulse and measure. $T_1$ and $T_2$ correlation measurements occur over a long acquisition time of $12~\text{hours}$ or more to acquire more than $1000$ `dual' event shots, which are then averaged. The resulting sets of data indicate a marked change in the expected qubit decay and dephasing populations after a `dual' event when compared to either `top' or `bottom' events.

The results of these experiments showed a statistically significant shift in the qubit’s expected $T_1$ population for a given time step, which agrees with the literature~\cite{McEwen2022,Harrington2025,Li2025}. However, we also measured a shift in the qubit’s expected $T_2$ population at a given time step. $T_1$ relaxes to its equilibrium value with a 1/e time constant of 38 $\mu\text{s}$. The $T_2$ effect relaxes on a similar timescale (1/e time constant of 25 $\mu\text{s}$).

\section{Discussion}

We found that our event statistics were consistent with the expected muon flux in our laboratory, which is $500'$ above sea level. When normalizing for the area of our detector substrates over 10 independent measurements, we measured $0.013 \pm 0.004$ dual events/s/cm$^2$ with an expected muon flux of $0.017$ events/s/cm$^2$~\cite{workman2022prog}. Single chip events were much more frequent, averaging approximately one event every ten seconds (0.067 single events/s/cm$^2$). These single-event MKID triggers are likely either caused by gammas being emitted from the chip packaging and various RF components~\cite{Loer2024} or misclassified dual events. Using our temperature calibration scheme, we extracted the normalized energy distributions of each event type (dual/single top/single bottom, see Fig.~\ref{fig:fig3}). Dual events appear to deposit less energy per detector on average, consistent with this interpretation.

We have observed a correlation between muon events and changes in a qubit’s coherence times. When an event is detected on both MKID arrays, a qubit chip in between the two detector arrays is more likely to experience impacted $T_1$ and $T_2$ times. Muons passing through the qubit substrate deposit energy which causes small bursts of localized heating. This heating creates quasiparticles in the qubit film, particularly at the Josephson junction where quasiparticle tunneling drastically reduces the qubit’s coherence. Qubit coherence times are reduced because of energy coupling between these quasiparticles and the qubit~\cite{Lenander2011}. Due to the variance in deposited muon energies (a result of absolute energy and trajectory), we are only observing the average effect from this quasiparticle tunneling in this data.

The observed effect in Fig.~\ref{fig:fig4}b is more difficult to fully explain since there are more fitting parameters in a Ramsey curve and we have set our observation point to a maximally frequency-sensitive readout time. The $T_2$ curve is affected by the qubit’s $T_1$ time and charge-sensitive noise processes such as parity jumps, charge dispersion, and charge offset could also play a role. Additionally, the qubit’s frequency can shift because of quasiparticle tunneling in tandem with the suppression of the superconducting gap in the presence of quasiparticles~\cite{Catelani2011}. Describing an exact model for the observed effect on $T_2$ is not possible given our current method of interrogation.

The $T_2$ data from Fig.~\ref{fig:fig4}b, which was taken over the course of $\sim$22 hours, exhibited a seemingly stochastic pattern in the average excited state population. While the target $P(1)$ was around $0.5$, the resulting average $P(1)$ ended up closer to $0.6$. We found that this discrepancy is caused by a drift in measured $T_2$ (see Appendix~\ref{app:drift}) but is averaged out over long timescales with data acquired at set intervals. Additionally, we are confident that this behavior did not affect the findings from Fig.~\ref{fig:fig4}b due to the reference curves (`top' and `bottom' MKID chips) representing an effectively random sampling and not demonstrating any shifts in the $T_2$ curve.

We also observed a smaller effect on $T_1$ and $T_2$ from top and bottom events (Figs.~\ref{fig:fig4}a and ~\ref{fig:fig4}b). For both cases, the top and bottom event curves are symmetric and have the same recovery period after the detected event as the dual-event data. We conclude that these deviations from the baseline measurement correspond to dual events that were falsely counted as single events. Taking a simplistic model that hypothesizes that qubit coherence times are only reduced by muon events, we use this data to estimate the fraction, $f$, of our single-event MKID triggers that should be attributed to misclassified dual events. Immediately after the detected event, $P(1)_{T_2}|_\text{single} = (1-f)+fP(1)_{T_2}|_\text{dual}$. The fraction is similar for the $T_1$ and $T_2$ datasets (18.4\% and 14.2\%, respectively). It is possible that a muon passing through the qubit could register on one of the MKID samples while either missing the other or failing to trigger due to mismatched detector sensitivities or some other unknown mechanism.  When we account for these misclassified events in our estimate of the muon flux from the $T_1$ ($T_2$) dataset, we measure 0.0154 (0.0141) events/s/cm$^2$, closer to the expected muon flux in our laboratory.

\section{Conclusion}

We have demonstrated a new platform for measuring correlated qubit errors with radiation events originating outside of the device. We observed both single and double chip blowouts, with the latter resulting in an increased likelihood of impact on a qubit’s decay ($T_1$) and phase coherence ($T_2$). These facets of qubit performance are likely affected due to muons heating the substrate and creating quasiparticles near the qubits’ Josephson junctions.

This platform enables close monitoring of radiation effects on qubit performance and can be used to study methods of improving muon shielding or mitigation techniques. However, these results have far-reaching implications for error correction codes since these events occur randomly at a rate of about 1 per minute. Understanding the exact way in which the qubit and readout resonator are affected could lead to targeted error-corrective measures for the case of cosmic radiation.

\begin{acknowledgments}
This work was performed under the auspices of the U.S. Department of Energy by Lawrence Livermore National Laboratory under Contract DE-AC52-07NA27344. LLNL-JRNL-2008718.
\end{acknowledgments}

\bibliography{MKID_references}

\appendix

\section{Measurement Setup and Pulse Sequencing}\label{app:setup}
\renewcommand{\thefigure}{A\arabic{figure}}
\setcounter{figure}{0}
\begin{figure*}[htbp]
    \includegraphics[width=\textwidth]{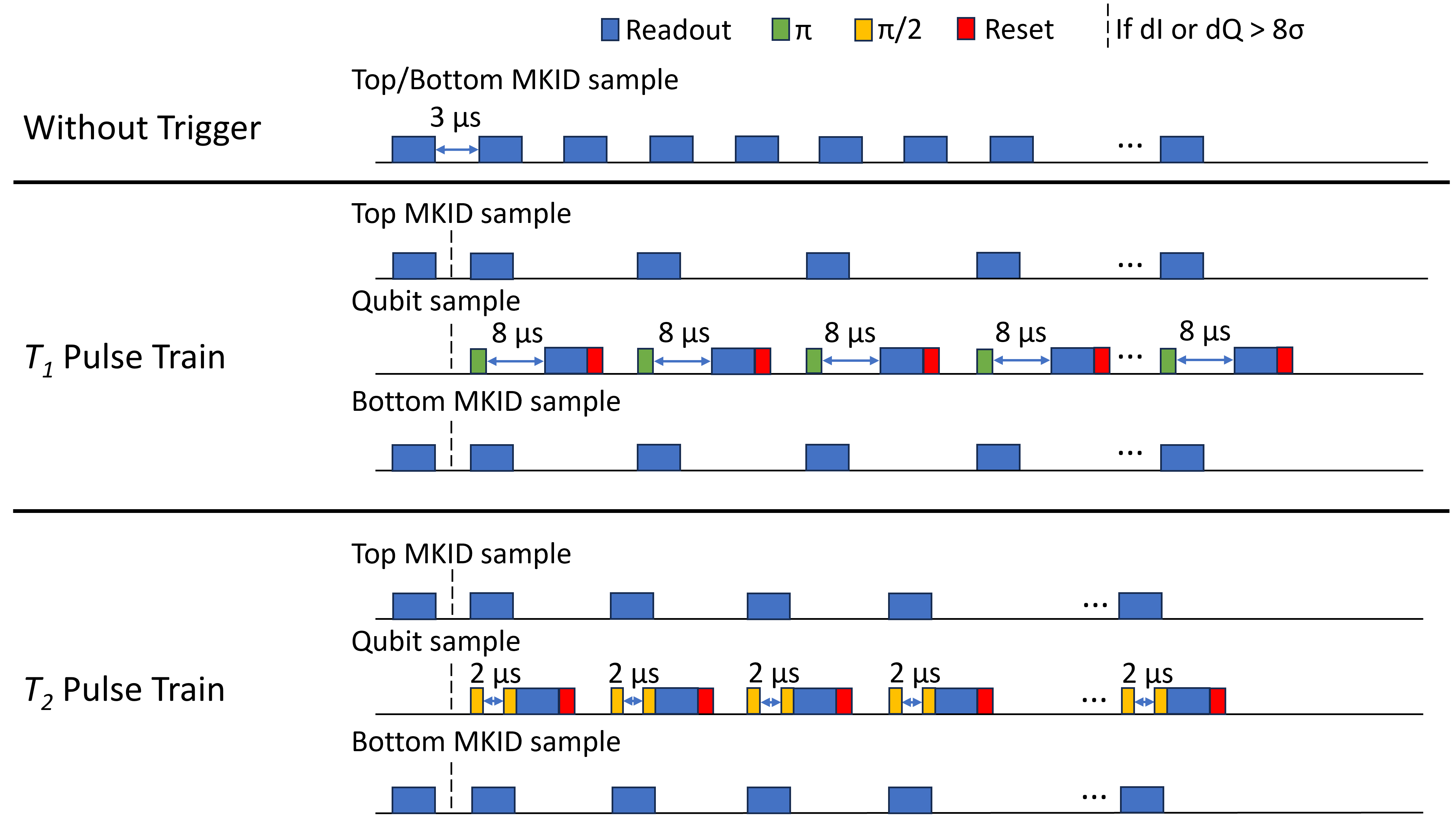}
    \caption{\textbf{Typical pulse configuration for MKID/qubit correlation measurements.} MKIDs are continuously measured with a 5 $\mu$s readout pulse in every 8 $\mu$s duty cycle. The delay between readout pulses is mostly due to buffering time on the FPGA. When an event is detected ($\Delta I$ and/or $\Delta Q$ $> 8\sigma$), the pulse train shown after the dashed line is executed fifty times (about 800 $\mu$s for $T_1$ and 560 $\mu$s for $T_2$) to obtain time-resolved event data, capturing both qubit coherence and the full MKID array response to the event. This approach ensures efficient data buffering and prevents processing overflows.}
    \label{fig:A1}
\end{figure*}

There are slight differences between experimental setups in statistics measurements (Fig.~\ref{fig:fig3}) and qubit correlation measurements (Fig.~\ref{fig:fig4}). The statistics measurements utilized two, three-by-three MKID arrays while qubit correlation measurements utilized two, five-by-five MKID arrays. In both sets of experiments, the MKID arrays were sandwiching a 6-qubit sample of aluminum junction, flux-tunable charge qubits with $T_1 \sim 25~\mu$s and $T_2 \sim 10~\mu$s. Their $E_J/E_C$ ranges from 35 to 50 and the qubit data presented in Fig.~\ref{fig:fig4} corresponds to a device with an $E_J/E_C$ of 36. The qubit measured in these experiments is located on the bottom left corner of the 6-qubit chip.

Measurement is performed in such a way as to prevent data and processing overflow on the FPGA hardware. This is done by using a data buffering technique in which we constantly measure the device but only trigger saves when the difference between initial and secondary measurements are above a set threshold. This threshold is determined by taking ten thousand background measurements for each resonator and finding the standard deviation, or $\sigma$, for each. We use $8\sigma$ as the threshold parameter because above this value, the number of events does not change by a significant amount indicating we are likely no longer counting false positives.

Both MKID samples and the qubit sample each occupy their own RF feedlines which are, in turn, each connected to their own RF control module. These three units are synchronized to operate in tandem, and each unit can measure up to nine signals at once. The MKID measurements run regardless of an event trigger but in the case of a detected event, both MKID and qubit time-resolved measurements begin. The pulse configuration and timing for a typical MKID/qubit event correlation measurement is shown in Figure~\ref{fig:A1}.

\section{Temperature Calibration}\label{app:calibration}
\renewcommand{\thefigure}{B\arabic{figure}}
\setcounter{figure}{0}

\begin{figure}[!htbp]
    \centering
    \includegraphics[width=1\linewidth]{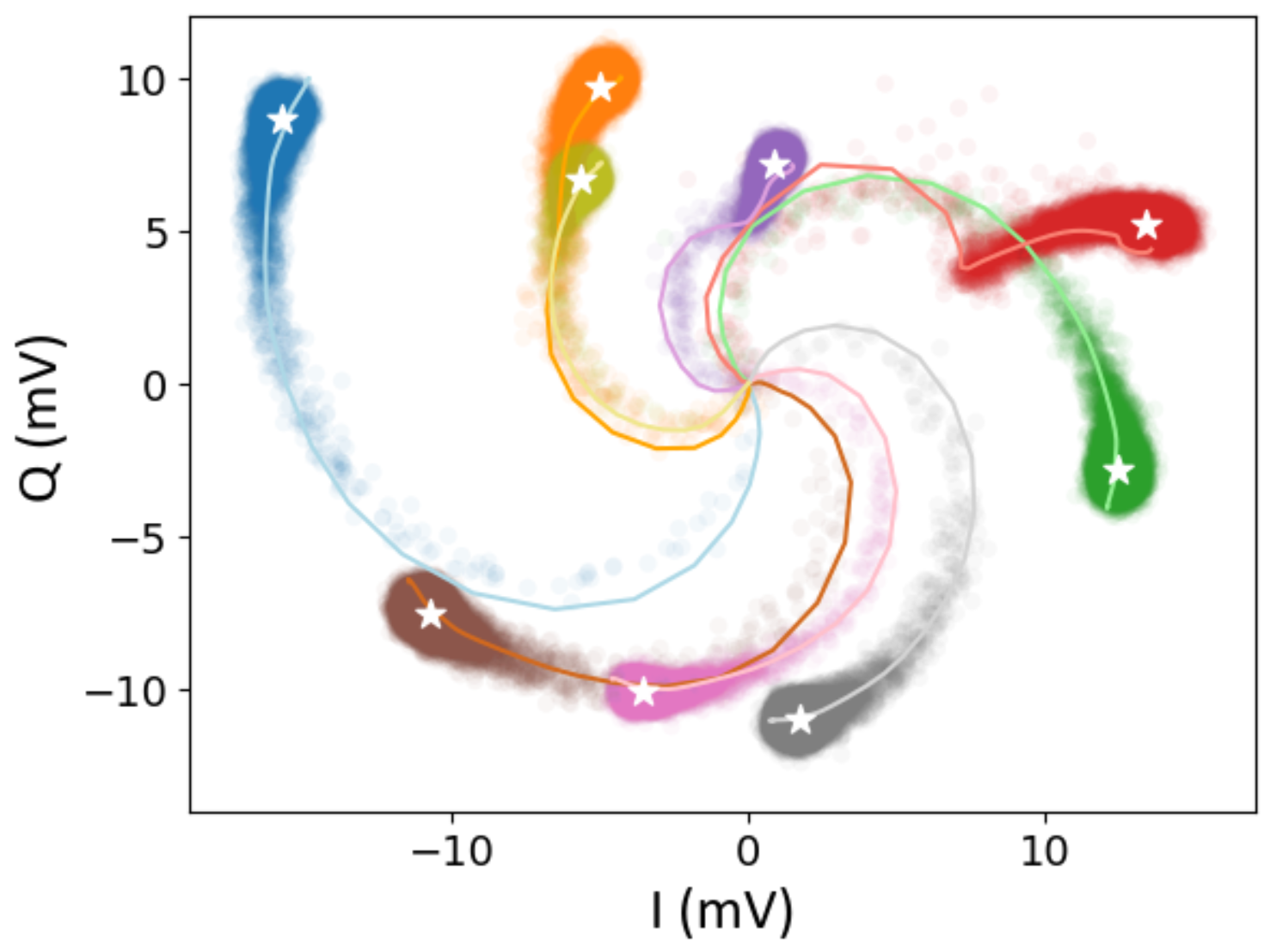}
    \caption{\textbf{Temperature calibration of MKID resonators.} For each MKID in the array, the IQ response was recorded as the dilution refrigerator temperature was swept from 10 mK to 300 mK. Spline fits to these data yield the the temperature calibration curves (solid lines; each color corresponds to a different MKID). Nonlinearities in some resonators manifest as kinks in the calibration curves. Stars mark the median IQ position for each resonator under standard operating conditions. Colored points represent triggered MKID responses to radiation events collected over a 12-hour period (as in Figure~\ref{fig:IQ}), overlaid to demonstrate that the calibration method accurately maps event-induced IQ shifts to effective temperatures.}
    \label{fig:B1}
\end{figure}

We calibrate the energy deposited during radiative events by assigning an effective temperature for each MKID response. We manually warm the dilution refrigerator and record each MKID’s average phase response at set temperature intervals. After acquiring this data, we then fit a 3D spline curve along the temperature axis to track the phase trajectory for each MKID. We can then define the temperature of newly acquired event data by correlating to the closest phase response in the fitted calibration data. Figure~\ref{fig:B1} demonstrates the calibrated IQ response of all detectors in the upper detector array as an example of the calibrated spline fits (dots connected by a solid line) overlaid with actual event detection data. The stars indicate the median IQ coordinate for each MKID.

After spline-fitting to the calibration data, we assigned an effective temperature for any newly acquired data by calculating the minimized distance to the spline curve. Whichever temperature is recorded at this location in IQ space becomes assigned to this individual data point and these assignments are then recorded. We calculate the minimal distance by taking the absolute value of the minimum of the difference between the temperature calibration curve coordinates and the data point coordinates we are trying to fit. The result gives us the closest possible value along the calibration curve for a given event data point.

The MKID data tended to accumulate phase drift over time which would manifest as a slow rotation about the origin of the IQ plane. We can set the calibration data as the original reference frame and rotate all subsequent accumulated data into this frame using a rotational transformation in polar coordinates. The operation for this rotation is given by

\begin{equation}
    C' = R C
\end{equation}

\noindent
where $R$ is the rotation matrix applied to $C$, the generalized median coordinate of the individual resonator data being transformed. $R$ takes the form of a $2 \times 2$ matrix

\begin{equation}
    R = \begin{pmatrix}
        \cos\theta & -\sin\theta \\
        \sin\theta & \cos\theta
    \end{pmatrix}
\end{equation}

\noindent
which we apply to $C$ in the form of a $2 \times 1$ matrix

\begin{equation}
    C = \begin{pmatrix}
        I \\
        Q
    \end{pmatrix}
\end{equation}

\noindent
thereby leaving us with a new coordinate in the form of another $2 \times 1$ matrix

\begin{equation}
    C' = \begin{pmatrix}
        I' \\
        Q'
    \end{pmatrix}
    .
\end{equation}

We found that applying this transformation to all the resonators’ data for a given run works well to align the original calibration data with the newly acquired phase-shifted data. It is possible for an event to behave differently than in our temperature calibration data due to the difference between equilibrium and non-equilibrium quasiparticle recombination. However, we have observed that recorded event data maps well to the temperature-calibrated spline fits.

\section{Investigating $T_2$ Measurement Drift}\label{app:drift}
\renewcommand{\thefigure}{C\arabic{figure}}
\setcounter{figure}{0}

\begin{figure}[!htbp]
    \centering
    \includegraphics[width=1\linewidth]{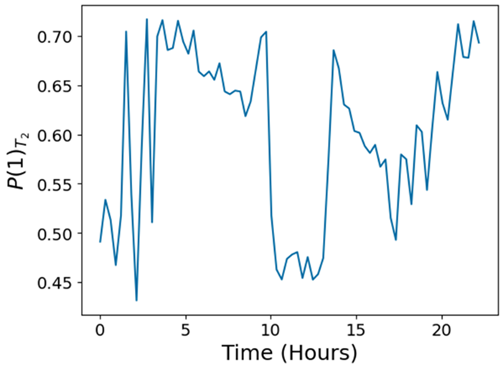}
    \caption{\textbf{Average single-shot $T_2$ measurement during a 22-hour acquisition.} Significant fluctuations in this measurement were observed throughout this period, spanning the full range of average signals detected in the triggered coherence time dataset (see Fig.~\ref{fig:fig4}). These variations may result from multiple underlying mechanisms, including quasiparticle-induced frequency shifts, energy decay, and charge noise processes such as charge dispersion and offset. The readout time was deliberately chosen to maximize sensitivity to frequency shifts, thereby amplifying the visibility of these noise sources. Importantly, the presence of these fluctuations does not affect the main conclusions, as untriggered coherence changes arise from mechanisms distinct from those associated with triggered events.}
    \label{fig:C1}
\end{figure}

The average single shot $T_2$ measurement (instead of specifically after events) reveals a relatively large drift over time (Figure~\ref{fig:C1}). The qubit in question has an $E_J/E_C$ of 36, so charge noise is a likely culprit. We used a qubit with this value of $E_J/E_C$ to investigate whether muon events were specifically responsible for any observed charge noise, but the results did not point to any causal relationship.

\begin{figure}[htbp]
    \centering
    \includegraphics[width=1\linewidth]{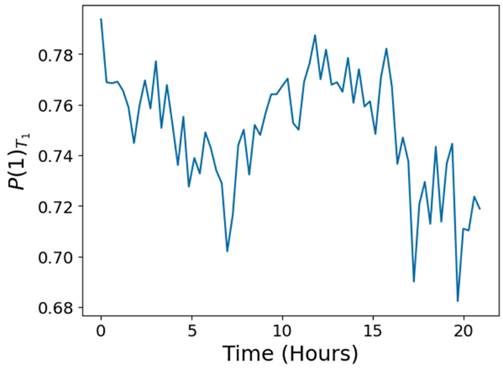}
    \caption{\textbf{Average single-shot $T_1$ measurement during a 22-hour acquisition.} The average excited state population slowly fluctuates by about 10\% over the full acquisition time, consistent with previous observations for this qubit system and likely explained by two-level system dynamics (see text). This drift is substantially smaller than the $\sim$35\% fluctuations observed in the $T_2$  measurement dataset. The relative stability of this measurement confirms that neither  qubit state preparation nor energy relaxation are responsible for the fluctuations observed in   $T_2$.}
    \label{fig:C2}
\end{figure}

In order to verify that the noise processes observed in Figure~\ref{fig:C1} did not affect qubit state preparation, we looked at an analogous dataset taken after single shot $T_1$ experiments (Figure~\ref{fig:C2}). While the data does exhibit some drift in the average excited state population, it is within expectations for this extended duration. More importantly, the large jumps we see in Figure~\ref{fig:C1} do not appear in Figure~\ref{fig:C2}. These datasets were taken at different times, but given the frequency of the jumps, we would expect to see something similar in $T_1$ experiments if the qubit state was affected. The drift seen in Figure~\ref{fig:C2} is likely caused by two-level systems (TLS) which are a common occurrence in superconducting qubit systems. These resonant systems fluctuate in frequency such that they enter the qubit’s bandwidth and absorb energy at different rates. The effect here is relatively small so this contribution does not significantly impact our results from Fig.~\ref{fig:fig4}a.

\begin{figure}[htbp]
    \centering
    \includegraphics[width=1\linewidth]{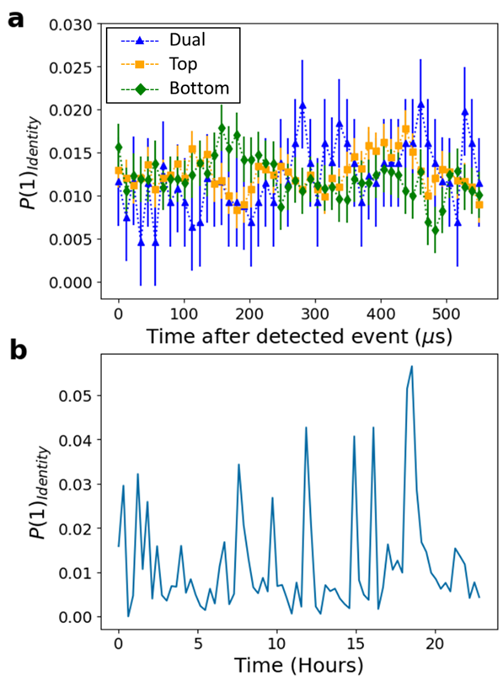}
    \caption{\textbf{Single-shot ground-state preparation experiment.} (a)  The mean expectation value of single-shot ground state preparation experiments is independent of event detection, as indicated by data markers (blue triangles: dual-event; orange squares: top-only; green diamonds: bottom-only). (b) The ground state population remains stable over the 22-hour acquisition period, typically exceeding 99\%. Occasional spikes in excited state population are observed, but these do not correlate with the fluctuations in the  $T_2$ measurement.}
    \label{fig:C3}
\end{figure}

Lastly, we investigated the ground state population of the qubit over a similar time duration to that of Figures~\ref{fig:C1} and~\ref{fig:C2}. We measured the ground state of the qubit immediately after a radiation event over a period of approximately 22 hours and analyzed the data in the same way as in Figure 5 (Figure~\ref{fig:C3}a). The resulting plot demonstrates the average excited state population of the qubit up to approximately 500 $\mu$s after three classes of events were detected. We found that there is no discernible excitation to the qubit immediately following any of the three event classes, which is consistent with the literature~\cite{McEwen2022,Harrington2025}. These results also indicate that we have a stable qubit ground state population of about 99\% which reinforces the notion that we have excellent control of the qubit state during $T_1$ and $T_2$ experiments.

We performed the same analysis on this data as in Figures~\ref{fig:C1} and~\ref{fig:C2} (Figure~\ref{fig:C3}b). Results indicate a stable qubit ground state population, often greater than 99\%, with occasional excursions of a few percent every few hours. These small spikes in the average excited state population are short-lived and do not appear to correlate with the observed drift in Figure~\ref{fig:C1}. There is no correlation between these spikes in excited state population and detected radiation events because there are approximately the same number of events in each data point represented in Figure~\ref{fig:C3}b. Furthermore, each data point represents about 18 minutes of data acquisition which is orders of magnitude longer than the quasiparticle recombination time for a typical radiation event.

Combining these results, we conclude that the findings in Fig.~\ref{fig:fig4} are strictly a result of increased quasiparticle density in the qubit film resulting from radiation interacting with the qubit substrate. Whatever the cause for the $T_2$ measurement drift, our results demonstrate a statistically significant shift in the qubit’s $T_1$ and $T_2$ response immediately after a radiation detection on both MKID chips simultaneously.

\end{document}